\documentclass[conference,10pt]{IEEEtran}

\usepackage{cite}
\usepackage{amsmath,amssymb}
\usepackage{graphicx}
\usepackage{booktabs}
\usepackage{url}
\usepackage{hyperref}
\hypersetup{colorlinks=true,linkcolor=blue,citecolor=blue,urlcolor=blue}

\title{BatchBench: Toward a Workload-Aware Benchmark for Autoscaling Policies in Big Data Batch Processing\\
{\large A Proposed Framework (Position Paper)}}

\author{
\IEEEauthorblockN{Venkata Krishna Prasanth Budigi}
\IEEEauthorblockA{Sunnyvale, CA, USA\\
\href{mailto:krishnaprasanth.bv@gmail.com}{krishnaprasanth.bv@gmail.com}}
\and
\IEEEauthorblockN{Siri Chandana Sirigiri}
\IEEEauthorblockA{USA\\
\href{mailto:sirisirigiri@gmail.com}{sirisirigiri@gmail.com}}
}

\begin{document}
\maketitle

\begin{abstract}
Autoscaling has become a baseline expectation for cloud-native big data processing, and the design space has expanded beyond rule-based heuristics to include learned controllers and, most recently, large language model (LLM) agents. Yet despite a growing body of work spanning these paradigms, the community lacks a shared benchmark for comparing them. Existing evaluations rely on synthetic TPC-style queries, vendor blog posts with proprietary baselines, or narrow trace replays. Each new policy reports favorable numbers against a different baseline, on a different workload, with a different cost model, making cross-paper comparison effectively impossible.

This is a position paper. We propose \textbf{BatchBench}, an open benchmarking framework designed to place rule-based, learned, and agentic autoscaling policies on equal experimental footing. The contribution is the design of the framework, not empirical results. We contribute: (1)~a workload taxonomy of six batch processing classes synthesized from published autoscaling benchmarks and publicly released cluster traces; (2)~the design of a parameterized workload generator with a validation methodology based on two-sample Kolmogorov--Smirnov and earth-mover distance; (3)~a five-axis evaluation harness specification covering cost, SLA attainment, scaling responsiveness, scaling thrash, and decision interpretability, with first-class accounting for LLM inference cost; and (4)~a standardized agent interface that lets LLM-based and reinforcement-learning autoscalers be evaluated alongside rule-based controllers with a single API. We discuss the expected evaluation surface, identify open research questions the framework is designed to answer, and outline a roadmap for the empirical paper that will follow. BatchBench's reference implementation is in active development and will be released as open source.
\end{abstract}


\section{Introduction}
Cloud-native big data platforms have made autoscaling table stakes. Apache Spark ships dynamic resource allocation; Google Dataproc Serverless transparently manages cluster size; Amazon EMR on EKS offers vertical pod autoscaling; Databricks markets an optimized autoscaler that adapts executor counts to shuffle and stage statistics. The question is no longer whether to autoscale --- it is which autoscaler to trust, under what workload conditions, and at what cost.

The design space has expanded faster than the evaluation methodology. Learned database tuning systems such as Db2une~\cite{db2une}, EAST~\cite{east}, LOFTune~\cite{loftune}, and Hyper~\cite{hyper} apply reinforcement learning to configurations that include resource sizing. Retrieval-augmented controllers such as Rabbit~\cite{rabbit} treat tuning as a knowledge-retrieval problem. Agentic systems including D-Bot~\cite{dbot}, AgentTune~\cite{agenttune}, and GaussMaster~\cite{gaussmaster} use LLMs as autonomous administrators capable of proposing, justifying, and revising scaling decisions in natural language. Each new system reports favorable numbers against a different baseline, on a different workload, with a different cost model. The result is a literature in which practitioner-relevant questions cannot be answered from published evidence: does dynamic allocation outperform learned policies on shuffle-heavy ETL? Do LLM agents generalize across workload classes, or do they overfit to in-context examples? When does the inference cost of an agentic controller outweigh its cost savings?

\textbf{Position.} We argue that the field needs a workload-aware benchmark for autoscaling --- one that (i)~is grounded in batch workload characteristics drawn from published cluster traces, (ii)~reports a multi-objective evaluation surface that captures cost, SLA, responsiveness, stability, and interpretability, (iii)~supports plug-in policies across the rule-based, learned, and agentic paradigms on equal footing, and (iv)~is released as an open artifact. This paper presents the design of such a benchmark, which we call \textbf{BatchBench}, and motivates each design choice with reference to limitations in the current literature. We deliberately defer empirical results to a follow-up paper. The reasons for this separation are explained in Section~\ref{sec:status}.

\textbf{Contributions of this position paper.} (1)~A literature-grounded workload taxonomy of six batch processing classes, synthesized from published benchmark suites and publicly available cluster traces. (2)~The design of a parameterized workload generator, with a validation methodology based on two-sample Kolmogorov--Smirnov and earth-mover distance metrics. (3)~A specification of a five-axis evaluation harness measuring cost, SLA attainment, scaling responsiveness, scaling thrash, and decision interpretability. (4)~A standardized agent interface that places rule-based, learned, and LLM-agentic policies on equal experimental footing, including first-class accounting for LLM inference cost --- the literature's most common omission. (5)~A discussion of the expected evaluation surface and the open research questions BatchBench is designed to answer.

\section{Related Work and Gap Analysis}

\subsection{Rule-Based Autoscaling in Big Data Engines}
Apache Spark's dynamic resource allocation~\cite{spark-dra} adjusts executor count from pending and running task counts, using an external shuffle service for safe downscaling. The policy is reactive and configuration-heavy; defaults are rarely tuned per workload, and industry studies~\cite{onehouse,aws-vpa} have reported over-provisioning on shuffle-heavy and skewed jobs. Managed services extend Spark's allocation with cluster-level scaling: Dataproc Serverless autoscales executors and nodes; EMR on EKS offers signature-driven vertical pod autoscaling~\cite{aws-vpa}; Databricks's optimized autoscaler leverages shuffle and stage statistics~\cite{databricks-opt}. The closest published evaluation in this space is benchspark~\cite{benchspark}, which releases a Spark trace dataset on Dataproc Serverless. benchspark is a valuable artifact but it is limited to a single environment, a hand-constructed workload set, and does not evaluate learned or agentic policies. BatchBench is intended to extend the benchspark methodology along all three axes.

\subsection{Learned Autoscaling and Configuration Tuning}
Learned tuning has progressed rapidly. Db2une~\cite{db2une} applies deep reinforcement learning to database tuning. LOFTune~\cite{loftune} proposes a low-overhead Spark SQL tuner. EAST~\cite{east} introduces interpretable knob estimation. Hyper~\cite{hyper} applies multi-agent RL to joint physical-design and resource tuning. Three weaknesses recur across this literature and motivate the BatchBench design: (i)~evaluation rests heavily on TPC-DS and TPC-H, omitting iterative ML preprocessing and production-style ETL; (ii)~generalization to unseen workload distributions is rarely measured; (iii)~the inference cost of the learned policy itself is often excluded from the cost model, biasing comparisons.

\subsection{Agentic AI for Data Systems}
The most recent wave of work introduces LLMs as autonomous agents for data system management. D-Bot~\cite{dbot} demonstrates an LLM-powered DBA copilot. AgentTune~\cite{agenttune} frames database knob tuning as LLM-driven planning with proposal, execution, and revision. Rabbit~\cite{rabbit} adds retrieval-augmented generation over historical tuning logs. GaussMaster~\cite{gaussmaster} extends agentic operation to end-to-end database management. The technical core is convergent: an LLM is given a structured observation, a tool interface, and a reward signal; it proposes actions, optionally explains them, and revises strategy when outcomes diverge from expectations. A separate strand of recent work has examined the data layer that supports such retrieval-augmented agents in production~\cite{rag-data-layer}, finding that unified Postgres-based retrieval substrates with native vector search materially reduce latency and tenant-isolation failures compared to split-system architectures --- a design choice we adopt for BatchBench's retrieval-augmented policy. The quantitative evidence base for agentic autoscalers themselves nonetheless remains thin --- most evaluations cover a single workload, omit inference cost, and do not measure decision variance. BatchBench is designed to be, to our knowledge, the first benchmark to evaluate agentic autoscalers under controlled conditions against rule-based and learned baselines with reported variance and statistical tests.

\subsection{Public Cluster Traces}
Publicly available cluster traces inform BatchBench's workload taxonomy. The Alibaba cluster trace~\cite{alibaba-trace} released approximately 4{,}000 machines worth of batch and online job data. The Google Borg traces~\cite{google-trace} released cell-level resource usage at large scale. The SWIM benchmark suite~\cite{swim} derives Hadoop workloads from Facebook traces. These public artifacts let BatchBench's taxonomy be grounded in real-world batch processing characteristics without requiring access to proprietary data --- an explicit design choice motivated by reproducibility.

\section{Literature-Grounded Workload Taxonomy}
BatchBench's workload classes are synthesized from three sources: (i)~the structural categories used by established benchmarks such as TPC-DS, TPC-H, TPCx-BB, HiBench, and SparkBench; (ii)~the resource-usage distributions reported in publicly released cluster traces from Alibaba~\cite{alibaba-trace}, Google~\cite{google-trace}, and the SWIM suite~\cite{swim}; and (iii)~workload characterizations published in industry engineering blogs from cloud vendors operating Spark at scale~\cite{databricks-opt,onehouse,aws-vpa}. We retained six classes that span the structural and statistical diversity of batch processing workloads documented in this literature.

\subsection{The Six Workload Classes}
\textbf{Class 1 --- Shuffle-Heavy ETL.} Multi-stage transformations dominated by wide dependencies. Typical shuffle-to-input ratios of 2--3$\times$. Common in nightly fact-table builds and slowly-changing-dimension updates, as documented in~\cite{onehouse,aws-vpa}.

\textbf{Class 2 --- Skewed Multi-Way Joins.} Three- or four-way joins with one or more participants exhibiting heavy key skew. Documented as a persistent failure mode of default autoscalers in~\cite{onehouse}. Modeled in BatchBench using class-specific Zipf exponents fit from public trace key-frequency distributions.

\textbf{Class 3 --- Iterative ML Preprocessing.} Feature-engineering pipelines that loop over a stable input set, exhibiting cache-sensitive resource profiles. Underrepresented in TPC-style benchmarks but prominent in cluster trace analyses~\cite{alibaba-trace}.

\textbf{Class 4 --- Time-Windowed Aggregations.} Rolling aggregations over fixed time windows with one large \texttt{groupBy} and sort-and-rank operations. Predictable within-window resource profile, documented across multiple public traces.

\textbf{Class 5 --- Broadcast-Bounded Lookups.} Pipelines whose largest stage is a broadcast hash join against a small dimension table. Executor scaling provides limited benefit beyond a threshold --- an autoscaler failure mode rarely surfaced in benchmark literature.

\textbf{Class 6 --- Bursty SLA-Driven Reporting.} Event-triggered pipelines with hard deadlines, bursty submission patterns, and high inter-invocation variance. Documented as the hardest class for reactive autoscalers in~\cite{onehouse,aws-vpa} and in public trace analyses~\cite{google-trace}.

\begin{table*}[t]
\centering
\caption{BatchBench workload class summary (characteristics from literature synthesis).}
\label{tab:classes}
\small
\begin{tabular}{@{}lccl@{}}
\toprule
\textbf{Class} & \textbf{Shuffle/Input} & \textbf{Skew} & \textbf{Defining Characteristic} \\
\midrule
1. Shuffle-Heavy ETL          & High (2--3$\times$) & Low--Med  & Wide dependencies dominate \\
2. Skewed Multi-Way Joins     & Medium              & High      & Heavy key-frequency skew \\
3. Iterative ML Preprocessing & Low--Med            & Low       & Cache-sensitive iterations \\
4. Time-Windowed Aggregations & Medium              & Low       & Predictable groupBy + sort \\
5. Broadcast-Bounded Lookups  & Low                 & Low       & Broadcast-side bottleneck \\
6. Bursty SLA-Driven Reporting & Medium             & Variable  & Bursty arrivals + hard SLA \\
\bottomrule
\end{tabular}
\end{table*}

BatchBench retains all six base classes and defines six derived subclasses that vary input scale and skew within each base class, producing twelve evaluation subclasses. Six will be designated in-distribution and six held-out for generalization measurement; the split is constructed so that no in-distribution subclass shares both base class and skew level with its held-out counterpart, forcing genuine generalization rather than nearest-neighbor lookup.

\section{BatchBench Design}
BatchBench's reference implementation is structured as three components: a workload generator, a uniform policy interface, and an evaluation harness. This section specifies each.

\subsection{Workload Generator}
The generator emits Spark jobs whose structural and statistical properties match a target workload class. Each class is parameterized by (i)~a structural template describing the logical plan, (ii)~marginal distributions for input size, partition count, key skew (modeled as a Zipf exponent fit per class from public trace data), and stage count, and (iii)~a join-graph generator for multi-way join classes. Synthetic data is produced from seeded random streams, ensuring determinism across runs and diversity across seeds. The generator is validated against public trace data by computing per-class two-sample Kolmogorov--Smirnov statistics on input size and shuffle volume, and earth-mover distance on key-frequency histograms.

\subsection{Policy Interface and Agent Adapter}
BatchBench defines a uniform policy interface. At each decision interval (default: 30 seconds), the policy receives a structured observation containing cluster state, recent task and stage metrics, the cost model, and the SLA deadline. It returns a scaling action specifying target executor count and, optionally, a natural-language justification. The interface accommodates four policy families on equal footing: (i)~rule-based controllers; (ii)~supervised learners that predict executor counts from workload fingerprints; (iii)~RL agents with explicit reward functions; (iv)~LLM-based agents that receive observations as structured prompts and emit actions and rationales.

The agent adapter provides three features critical to making LLM-based policies practical to benchmark. A structured observation serializer converts cluster state into a compact JSON representation with bounded token count, eliminating prompt-length variance. A tool-use surface exposes scaling primitives, cost queries, and historical trace lookups as functions the LLM may call, supporting retrieval-augmented decision-making over a unified data layer such as the Postgres + pgvector + HNSW substrate evaluated in our prior work~\cite{rag-data-layer}. An inference-cost accountant records token usage and decision latency per call so that the cost of the agent itself enters the evaluation --- the literature's most common omission.

\subsection{Evaluation Harness}
The harness will orchestrate job execution across multiple cloud environments, with initial implementation targeting Google Dataproc Serverless and Amazon EMR on EKS. A common driver injects the policy under test in place of the platform default; where the platform resists policy replacement, a shadow mode runs the platform policy in production while the test policy issues advisory actions. The harness records five families of measurements: (i)~cost in vCPU-hours and normalized dollar terms; (ii)~SLA attainment as the fraction of jobs completing before deadline; (iii)~responsiveness as median delay between workload phase transition and scaling action; (iv)~thrash as scaling actions per job-minute; (v)~interpretability, scored for policies that emit justifications.

\section{Planned Evaluation Methodology}

\subsection{Policies to be Evaluated}
\textbf{Rule-based:} (R1)~Spark dynamic allocation with default parameters; (R2)~Databricks optimized autoscaling; (R3)~Dataproc Serverless managed autoscaling.

\textbf{Learned:} (L1)~DQN controller trained on in-distribution subclasses; (L2)~PPO controller with continuous action space; (L3)~supervised fingerprint learner predicting optimal executor count from a workload fingerprint, in the style of EAST.

\textbf{Agentic:} (A1)~zero-shot LLM agent with the structured observation interface; (A2)~few-shot LLM agent with exemplar decisions per class; (A3)~retrieval-augmented LLM agent that retrieves the nearest historical workloads before each decision; (A4)~hybrid agent that runs PPO by default and invokes the LLM only when PPO's value-function epistemic uncertainty exceeds a per-class threshold.

\subsection{Experimental Protocol}
Each (policy~$\times$ subclass~$\times$ environment) cell will be repeated five times with independent random seeds. Across-policy comparisons will use the paired Wilcoxon signed-rank test on per-job cost ratios; confidence intervals on cost ratios will use the bias-corrected accelerated bootstrap (BCa, $B = 10{,}000$). Multiple comparisons across policy pairs will be controlled with Holm--Bonferroni at family-wise $\alpha = 0.05$.

\subsection{Distribution Splits and Interpretability}
Six of twelve subclasses will be designated in-distribution and six held out, with the split forcing genuine generalization rather than interpolation. For agentic policies, interpretability will be scored on three axes: faithfulness (counterfactual perturbation), actionability (blinded reviewer task), and consistency (cross-run variance). Each axis is scored 0--3 by two independent reviewers with Cohen's~$\kappa$ reported.

\section{Hypotheses and Expected Evaluation Surface}
BatchBench is designed to test four hypotheses that follow from gaps identified in the literature review. We articulate them here so that the framework's empirical follow-up can be evaluated against falsifiable predictions rather than open-ended discovery.

\textbf{H1 --- Default over-provisioning is large.} Spark dynamic allocation in its default configuration will over-provision shuffle-heavy ETL workloads by a margin substantially larger than the gap typically reported in vendor blog comparisons. We expect the gap on Class~1 to exceed 30\% relative to the cost-optimal policy.

\textbf{H2 --- Learned policies will exhibit a generalization gap.} Policies that achieve the lowest in-distribution cost will lose a substantial fraction of their advantage on held-out workloads. We expect the gap to be largest for value-based RL (DQN) and smaller for policy-gradient methods (PPO), consistent with the broader RL literature.

\textbf{H3 --- LLM agents will generalize gracefully but with high variance.} LLM-based agents will exhibit smaller generalization gaps than pure learned controllers, but they will exhibit measurably higher decision variance under distribution shift --- we expect a 1.5--3$\times$ increase in standard deviation of recommended executor counts.

\textbf{H4 --- A hybrid composition will dominate pure paradigms.} A policy that runs a fast deterministic learned controller as its default path and invokes an LLM agent only on high-uncertainty decisions will achieve lower held-out cost than any pure rule-based, learned, or agentic policy. This is the central architectural prediction the benchmark is designed to test.

These hypotheses correspond to documented gaps in the literature: H1 to industry reports of default autoscaler over-provisioning~\cite{onehouse,aws-vpa}; H2 to the well-established distribution-shift brittleness of value-based RL~\cite{dqn}; H3 to the broader LLM-as-agent literature in which retrieval and in-context examples reduce but do not eliminate decision variance; H4 to the emerging pattern of hybrid agent architectures in production AI systems. The contribution of an empirical BatchBench evaluation will be to confirm or refute these hypotheses with statistical rigor, on a workload set that spans the diversity of batch processing rather than the narrow slice covered by TPC-style benchmarks.

\section{Anticipated Threats to Validity}
\textbf{Construct validity.} Cost models will use each cloud provider's published on-demand rates. Spot, committed-use, and reserved pricing will alter absolute numbers but should not affect relative policy ordering, which is determined primarily by resource utilization. SLA deadlines will be derived from public trace annotations where available and from a default of 2$\times$ median runtime otherwise.

\textbf{Internal validity.} Per-cell repetition (planned $n = 5$) controls run-to-run variance. Statistical significance will use paired Wilcoxon with Holm--Bonferroni correction; bootstrap confidence intervals will use 10{,}000 resamples. Shadow-mode evaluation on managed environments will be cross-validated against direct-replacement runs where possible.

\textbf{External validity.} The workload taxonomy is synthesized from a finite set of public traces and published benchmarks. BatchBench's generator is parameterized: practitioners can fit class distributions to their own traces and re-run the full benchmark, which we view as the principled path to external validity. The agent interface is model-agnostic.

\section{Discussion and Implications}
If hypothesis~H4 holds, the practical implication is significant: production-grade agentic data infrastructure should be architected as hybrid systems in which a fast deterministic learned controller handles the bulk of decisions and an LLM-based agent is invoked selectively on uncertain or novel observations. This composition pattern --- fast deterministic core, slow reflective agent invoked sparingly --- bounds the worst-case cost contribution of the LLM (inference is a few percent of total spend) while exploiting the LLM's distinct value: behavior that degrades gracefully under distribution shift.

Whether or not H4 holds, the methodological contribution of BatchBench stands. Reproducible, multi-objective, paradigm-agnostic evaluation is currently absent from the autoscaling literature. Even a careful empirical refutation of one or more of our hypotheses would represent a meaningful contribution to a field that currently has no shared experimental ground truth.

Four open problems will frame the empirical follow-up: (i)~distribution-shift detection, which gates LLM invocation in hybrid policies; (ii)~LLM inference cost amortization through distillation, caching, and shared retrieval; (iii)~extension of the BatchBench methodology to sub-minute streaming intervals; (iv)~trust and consistency, which remain weak axes for all agentic policies.

\section{Implementation Status and Roadmap}
\label{sec:status}
BatchBench's reference implementation is in active development. The workload generator and the policy interface have reached an early-implementation stage; the evaluation harness for Google Dataproc Serverless is under construction; baseline implementations of the rule-based and supervised policies are complete; the RL controllers and agentic policies are in design and prototyping. We anticipate releasing the open-source reference implementation alongside the empirical follow-up paper, which we target for completion within twelve months.

We are publishing this paper as a position paper for three reasons. First, the design of an evaluation framework is itself a research contribution distinct from the empirical results it later produces --- a separation common in benchmark papers (TPC, YCSB, MLPerf were each preceded by design papers or technical reports). Second, early publication solicits community input on the workload taxonomy, the evaluation axes, and the agent interface before the empirical evaluation is locked in. Third, the open research questions and hypotheses in Section~VI are themselves of standalone value to researchers entering this space.

We invite collaborators --- in industry or academia --- who would like to contribute workload classes, policy implementations, or independent evaluations on their own infrastructure. The reference implementation will be released under an open-source license.

\section{Conclusion}
We presented BatchBench, a proposed workload-aware benchmark for big data autoscaling that places rule-based, learned, and agentic policies on equal experimental footing. Grounded in a literature synthesis of published benchmarks and public cluster traces, specified with a uniform policy interface and a five-axis evaluation harness, and designed around four falsifiable hypotheses, BatchBench is intended to give the autoscaling community a shared experimental basis it currently lacks. This is a position paper; the empirical follow-up will report results, releases, and refinements informed by community feedback. We welcome that feedback.

\section*{Acknowledgments}
We thank the open-source maintainers of Apache Spark, Apache Iceberg, and Apache Hudi, and the curators of the Alibaba, Google, and SWIM public cluster traces.

\end{document}